\documentclass[prb,twocolumn,groupedaddress]{revtex4}
\usepackage{graphicx}
\usepackage{pifont}
\usepackage{amssymb}
\usepackage{epsfig}
\usepackage{amssymb}
\usepackage{amsmath}
\usepackage{color}

\DeclareMathAlphabet{\mathitb}{OT1}{cmr}{bx}{sl}
\begin{document}

\renewcommand{\thefootnote}{\fnsymbol{footnote}}
\title{Anomalous Transport and Possible Phase Transition in Palladium Nanojunctions}
\author{Gavin D. Scott$^1$}
\email{gavin.scott@rice.edu}
\author{Juan J. Palacios$^2$}
\author{Douglas Natelson$^{1,3}$}

\affiliation{
$^1$Department of Physics and Astronomy, Rice University, 6100 Main St., Houston, TX 77005\\
$^2$Departamento de F\'{i}sica de la Materia Condensada, Universidad Aut\'{o}noma de Madrid, Campus de Cantoblanco, Madrid 28049 Spain\\
$^3$Department of Computer and Electrical Engineering, Rice University, 6100 Main St., Houston, TX 77005}
\date{\today}

\begin{abstract}

Many phenomena in condensed matter are thought to result from competition between different ordered phases.  Palladium is a paramagnetic metal close to both ferromagnetism and superconductivity, and is therefore a potentially interesting material to consider.  Nanoscale structuring of matter can modify relevant physical energy scales leading to effects such as locally modified magnetic interactions.  We present transport measurements in electromigrated palladium break junction devices showing the emergence at low temperatures of anomalous sharp features in the differential conductance.  These features appear symmetrically in applied bias and exhibit a temperature dependence of their characteristic voltages reminiscent of a mean field phase transition.  The systematic variation of these voltages with zero-bias conductance, together with density functional theory calculations illustrating the relationship between the magnetization of Pd and atomic coordination, suggest that the features may result from the onset of spontaneous magnetization in the nanojunction electrodes.  We propose that the characteristic conductance features are related to inelastic tunneling involving magnetic excitations. \\

\end{abstract}

\maketitle 

\textbf{Introduction}\

Many phenomena in modern condensed matter physics, including metal-insulator transitions, high temperature superconductivity, heavy fermions, multiferroic effects, and quantum criticality, are thought to originate from competition between different possible ordered electronic ground states.  Slightly shifting the balance between these orders via external fields, strain, disorder, or carrier concentration can tune a system between different phases.  Palladium's proximity to both ferromagnetism\cite{Fritsche1987,Taniyama1997,Lee1998,Sampedro2003} and
superconductivity\cite{Skoskiewicz1972,Stritzker1972,Stritzker1979} make it an interesting material to consider in this context.  Structuring matter on the nanometer scale is one means of modifying relevant physical energy scales, with nanoscale confinement already known to favor locally modified magnetic interactions in Pd.\cite{Taniyama1997,Delin2004,Ralph2008,Calvo2009}  Nanogap structures can be particularly useful in examining the relevant physics in such complicated systems.  Tunneling transport has been a valuable spectroscopic tool, with the tunneling conductance as a function of bias voltage typically interpreted as probing the convolution of the electronic densities of states (DOS) of the two electrodes.  In fixed nanogap structures, this has given insight into varied systems, including charge density wave compounds\cite{ONeill2006} and assorted superconductors.\cite{Schmidt2002,Ekino1996}

\begin{figure}[!b]
\begin{center}
\includegraphics[scale = .21]{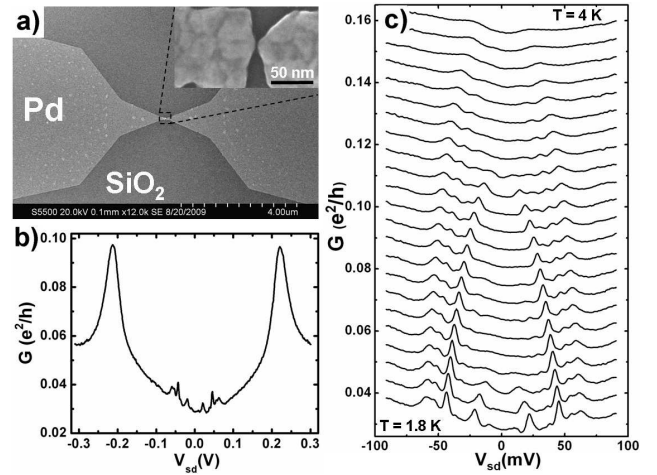}  
\end{center}
\hspace\fill \vspace{-9mm} \caption{(a) Scanning electron microscope (SEM) image of a Pd nanowire on SiO$_2$.  Inset: High resolution image after electromigration exhibiting a tunneling gap separation $<$ 2~nm and an absence of nanoparticles.  (b) $dI/dV$ \textit{vs.} $V_{\mathrm{sd}}$, at $T$ = 1.8~K, for representative sample \textbf{1}, which includes a pair of broad high bias peaks near $\pm$215 mV and a series of narrow features at lower bias, nearly symmetric about $V_{\mathrm{sd}}$ = 0 V.  (c) Waterfall plot of the lower bias region of sample \textbf{1}, showing the temperature evolution.  Temperature is increased from 1.8 K (bottom) to 4 K (top) in steps of 100 mK.  Traces are offset for clarity by 0.006 $e^2/h$.}
\label{figure1}
\end{figure}

In this article we report tunneling measurements in Pd-Pd tunnel
junctions fabricated by electromigration of planar constrictions on
oxidized Si substrates.  We find a characteristic pattern of features
in the differential conductance appearing at source-drain voltages
($V_{\mathrm{sd}}$) linearly related to the log of the zero-bias conductance
(ZBC $\equiv G(V_{\mathrm{sd}}=0)$).  These features rapidly decrease in bias with increasing temperature, with a temperature dependence reminiscent of a
mean-field transition.  Of particular interest is the contrast in
scale between the energy of an individual peak or dip ($eV_{\mathrm{sd}}$) and
the respective temperature at which it vanishes ($k_BT_{\mathrm{c}}$), which often
differ by $\sim$ 100$\times$.  These features are robust and appear to
be particular to Pd electrodes.  We discuss possible physical
mechanisms for these features, and suggest that
they are likely signatures of inelastic tunneling processes
associated with the onset of spontaneous magnetization in the electrode
tips.  While the details have not been definitely determined, calculations of the
expected magnetic moment in model Pd nanocontacts are consistent
with both the onset temperature of the conductance features and
the observed systematic variation of voltage scales with the ZBC.

\textbf{Results}\

Bowtie-shaped Pd nanowires are fabricated on a Si/SiO$_2$ substrate
(see Methods).  The devices are cleaned and subsequently
cooled to $\lesssim$ 2~K.  A tunneling gap is created in a nanowire by
employing the electromigration
technique (Fig.~\textbf{1}a).\cite{HPark1999}  Differential
conductance ($dI/dV$) is measured as a function of
$V_{\mathrm{sd}}$, and the conducting substrate is used as a back
gate by applying a separate bias, $V_g$.

When the electromigration process reduces the ZBC below $G_0$ (most
commonly between $0.30G_0$ and $0.001G_0$, where $G_0 = e^2/h$), a
characteristic series of peaks and dips is observed in the curve of
$dI/dV$ \textit{vs.} $V_{\mathrm{sd}}$ when measured at low temperatures.  These
differential conductance features (DCFs) are highly reproducible in
Pd.  Out of 111 clean Pd nanowire samples successfully electromigrated
to a measurable ZBC less than $G_0$, more than 75\% of devices
exhibited characteristic DCFs and their corresponding temperature
dependence.  This holds true for samples fabricated on both $n$- and
$p$-type Si substrates.  When the electromigrated nanojunctions have a
ZBC of $G(V_{\mathrm{sd}}=0) \geq G_0$, these features are entirely absent. These DCFs appear to be specific to Pd.  Experiments with both Ni-Ni and Au-Au tunnel junctions formed via analogous fabrication processes have shown no indications of similar DCFs.

\begin{figure}
\begin{center}
\includegraphics [scale = .22]{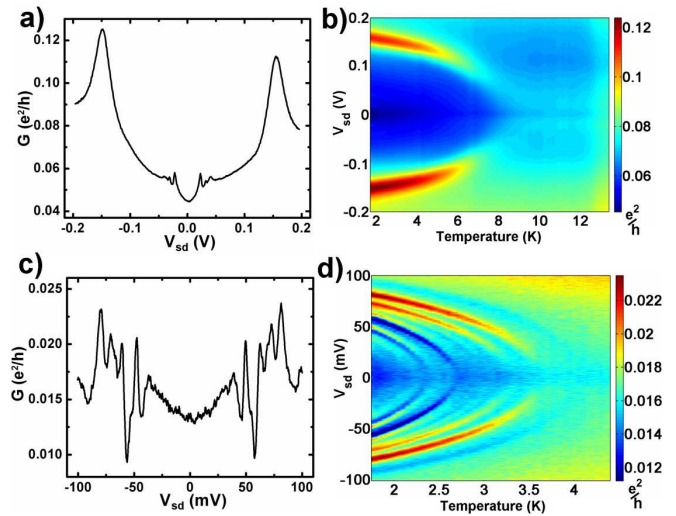} 
\end{center}
\hspace\fill \vspace{-9mm} \caption{Base temperature traces of $dI/dV$ \textit{vs.} $V_{\mathrm{sd}}$ and their corresponding temperature dependence.  (a) $dI/dV$ \textit{vs.} $V_{\mathrm{sd}}$ for device \textbf{2}.  (b) Colormap depicting temperature evolution of device \textbf{2}.  Conductance amplitude is indicated according to colorbar on the right.  The high bias peaks terminate near 9 K.  The lower bias features evolve in a similar fashion, but are not observable on this color scale.  (c) $dI/dV$ \textit{vs.} $V_{\mathrm{sd}}$ for device \textbf{3} showing only the features in the low bias regime.  Many peaks and dips are evident.  (d) Colormap of temperature evolution for device \textbf{3}.  Each feature in the spectrum of $dI/dV$ disappears at a different value of $T$, which we refer to as its critical temperature, $T_{\mathrm{c}}$.}
\label{figure2}
\end{figure}

\begin{figure*}[!t]
\begin{center}
\includegraphics [scale = .33]{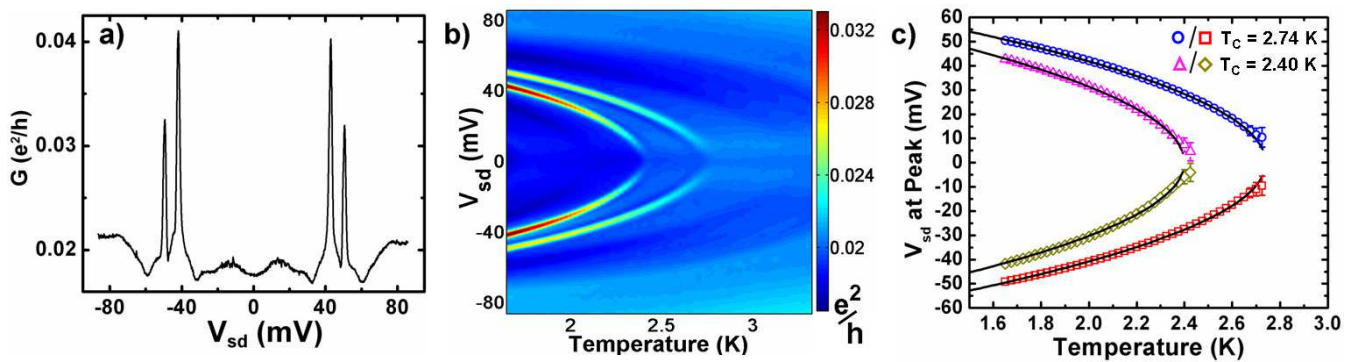}  
\end{center}
\hspace\fill \vspace{-9mm} \caption{Temperature dependence of the voltages at which DCFs appear is well characterized by equation (\ref{MeanField}).  (a) $dI/dV$ \textit{vs.} $V_{\mathrm{sd}}$ at T = 1.65 K for sample \textbf{4}.  (b) Colormap of sample \textbf{4} formed from traces of $dI/dV$ \textit{vs.} $V_{\mathrm{sd}}$ acquired at temperatures between 1.65 K and 3.325 K in steps of 25 mK.  (c) Plot of voltage at which the two primary conductance peaks (on both sides of $V_{\mathrm{sd}}$ = 0 V) appear versus temperature.  Solid lines are least squares fits to the data using equation (\ref{MeanField}).  Error bars represent uncertainty in peak position of Gaussian lineshape fit to each peak using a least squares method.}
\label{figure3}
\end{figure*}

When $T \sim$ 1.8~K we distinguish distinct types of DCFs associated
with two different bias regimes.  In the ``low bias'' region
(typically $|V_{\mathrm{sd}}| \lesssim$ 110~mV), a series of sharp features
appears on each side of $V_{\mathrm{sd}}$ = 0~V (Fig. \textbf{1}b $\&$ c).
From device to device, these peaks and dips occur with varying
amplitudes, lineshapes, and spacings; for a given device, the features
are clean and reproducible.  The ``high bias'' region (typically
$|V_{\mathrm{sd}}| \approx$ 150~mV - 400~mV) contains the second type of
feature, which is always manifested as a single pair of relatively broad
peaks, with one peak appearing on each side of $V_{\mathrm{sd}}$ = 0 V
(Fig. \textbf{1}b).  The amplitude of these two peaks is greater
than those in the low bias region.  While the pattern of all DCFs
appear symmetric about $V_{\mathrm{sd}}$ = 0~V, there can be an overall
asymmetry in conductance magnitudes due to a background contribution
(roughly linear in $V_{\mathrm{sd}}$) to $dI/dV$.  Such a background is commonly
observed in STM and break junction experiments.\cite{Nagaoka2002,Calvo2009}

The DCFs undergo a rapid decrease in voltage position and amplitude as
temperature is increased (Fig. \textbf{1}c).  At sufficiently high
temperatures all temperature-evolving features have disappeared, and
the resulting trace of $dI/dV$ appears weakly nonlinear, consistent
with what is expected for an ordinary metallic break junction device
with a small interelectrode gap.  A critical temperature, $T_{\mathrm{c}}$, can
be assigned to each DCF, defined as the temperature at which that
particular feature approaches $V_{\mathrm{sd}}$ = 0 V.  The energy of an
individual peak or dip cannot usually be tracked all the way to
$eV_{\mathrm{sd}}$ = 0 V as it invariably becomes difficult to resolve with
diminishing amplitude, resulting in the apparent ``smearing'' of the
colormaps in the vicinity of the $V_{\mathrm{sd}}$ = 0~V axis
(Figs. \textbf{2}, \textbf{3}, and \textbf{4}).  The energies
of the DCFs decrease with temperature in a manner that is well fit by
the equation:
\begin{equation}
eV_{\mathrm{sd}}(T) = A\left(1 - \frac{T}{T_C}\right)^\beta,
\label{MeanField}
\end{equation}
where $A$ is the voltage position of the peaks (multiplied by $e$) as $T \rightarrow$ 0~K and $\beta = \frac{1}{2}$, the classical exponent for a mean-field transition in a Landau theory.  The low bias DCFs of all Pd devices have $T_{\mathrm{c}} \lesssim$ 4.3~K and evolve with temperature as in
Equation~(\ref{MeanField}).  An example is shown in
Fig.~\textbf{3}, along with fits to Equation~(\ref{MeanField}).  The
energy of the high bias peaks also follows this form, but begins to
deviate at $T/T_{\mathrm{c}} \lesssim \frac{1}{2}$, where $T_{\mathrm{c}}$ is typically between 8-12 K (Fig.~\textbf{2}b and Supplementary Fig. \textbf{S5}).

In Figure \textbf{4} the voltages of representative peaks and dips, from a set of several dozen devices, are plotted as a function of the log of their corresponding ZBC when $T \approx$ 1.65 K.  The peaks associated with high bias regions are quite linear in log($G(V_{\mathrm{sd}}=0)$), indicating that the voltage position of these features decays exponentially with ZBC.  Fits to the high bias data indicate that these peaks will go to zero bias as $G(V_{\mathrm{sd}}=0) \rightarrow$ $G_0$.  This is consistent with our observation that DCFs are only present for samples with $G(V_{\mathrm{sd}}=0) < G_0$.  DCFs in the plot associated with the low bias regime appear anywhere within a bias range between $V_{\mathrm{sd}}$ = 0 V and an upper $|V_{\mathrm{sd}}|$ limit.  The edge of this envelope is also roughly linear in the log of $G(V_{\mathrm{sd}}=0)$.

\begin{figure}[b]
\begin{center}
\includegraphics [scale = .36]{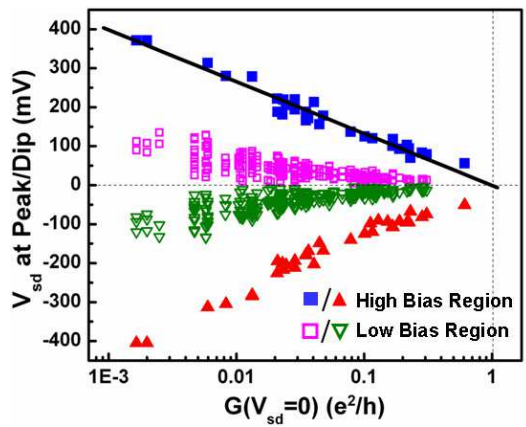}
\end{center}
\hspace\fill \vspace{-9mm} \caption{Exponential relationship between voltage position of DCFs and corresponding ZBC.  $V_{\mathrm{sd}}$ position of representative $dI/dV$ features at $T \approx$ 1.65 K for many junctions plotted as a function of the log of $G(V_{\mathrm{sd}}=0)$ for each sample.  Points pertaining to high bias peaks follow linear curve.  Black line is a linear fit to positions of the high bias peaks for $V_{\mathrm{sd}} >$ 0 V.  Points pertaining to low bias features appear between $V_{\mathrm{sd}}$ = 0 V and an envelope outlined by an approximately linear curve.}
\label{ZBCvsVsd}
\end{figure}


The ZBC of a tunnel junction with a vacuum gap is expected to decrease
exponentially with increasing interelectrode gap distance, $d$, according to
$G \propto \mathit{e}^{-{\alpha}\mathit{d}}$, where $\alpha \propto \sqrt{\phi}$, and $\phi$ is a typical barrier height related to the metal work function.  Since $V_{\mathrm{sd}} \propto$ log($G(V_{\mathrm{sd}}=0 V)$), a linear relationship exists between the $V_{\mathrm{sd}}$ positions of the DCFs and $d$.  This is further confirmed by observations in a single device when measurements of $dI/dV$ \textit{vs.} $V_{\mathrm{sd}}$ are performed after successive electromigrations: The ZBC is reduced and the DCFs are found at higher biases while the pattern of DCFs remains relatively unchanged (see Fig. \textbf{S2} in Supplementary Information). 

Conversion of the ZBC data in Fig.~\textbf{4} into a distance scale is extremely sensitive to both the conductance prefactor and $\alpha$, neither of which is well established in Pd.\cite{Minowa2005}  If desired, one could perform theoretical calculations based on density functional theory to obtain a quantitative estimate for $\alpha$ with particular tip geometries.  A typical value of $\alpha$ for vacuum tunneling between high work function metals is 2 \AA$^{-1}$.\cite{Wolf1985}  Using this value and assuming a prefactor for the conductance of 2$e^2/h$ when the two electrodes are in contact at the single atom level, we then estimate that $d$ ranges between 1 and 4 \AA.


We have also examined the effects of gate bias and magnetic field on
the junctions, as further explained in the Supplementary Information.  As
$V_g$ is increased in the positive (negative) direction, the ZBC
decreases (increases), and the DCFs move to higher (lower) energies,
increasing (decreasing) their associated $T_{\mathrm{c}}$'s (Fig. \textbf{S1}c).  Some DCFs appear to split at large negative $V_{\mathrm{g}}$.  There is no sign of
Coulomb stability diamonds, and these gate bias effects have the same
polarity for both $n$- and $p$-type substrates.  External magnetic
fields up to $\pm$9~T in a direction perpendicular to the plane of the
devices lead to no significant effects.  There is no Zeeman splitting
or broadening of peaks in $dI/dV$, and no hysteresis seen in
measurements of magnetoresistance (Fig. \textbf{S1}d).  There is a very slight
shift ($\sim 1 mV$) of DCFs to higher voltages at $\pm$9~T, but we
cannot rule out that this is due to slight field-dependent changes in
sample temperature.  SQUID measurements of co-deposited Pd films show
only paramagnetic response down to 2.0~K.


\textbf{Discussion}\
 
Selectively altering matter on the nanometer scale is one means of modifying relevant physical energy scales and can lead to collective electronic effects not necessarily present in bulk samples.  The anomalous transport data presented here, found in a chemically homogeneous system, are an example of such an effect.  We consider several possible explanations for the observed phenomena.
One concern is that the DCFs are extrinsic rather than an inherent
property of the Pd junctions.  For example, the presence of helium
exchange gas in the chamber raises the possibility that an adsorbed He
film could be the origin of the features.  This is unlikely however,
since the DCFs are also observed when no He exchange gas is used.
Also, with or without exchange gas, the high bias DCFs survive well
above 4.2~K.

Similarly, it is well established that hydrogen has an affinity for
Pd.\cite{Baykara2004}  Previous experiments involving tunneling
spectroscopy of metallic nanogaps in the presence of H$_2$
reveals low temperature $dI/dV$ features reminiscent of
our data due to transitions between coverage dependent
states\cite{Gupta2005} or vibrational modes\cite{Thijssen2006} of the
molecules.  The voltages of the DCFs we report are similar in scale to
the energy of phonons or vibrational modes related to adsorbed
molecules.  However, phonon mode energies do not have the strong
temperature dependence exhibited by these DCFs.  Furthermore, unlike
the H$_2$-containing junctions, our devices are stable over time (3-4
days), do not exhibit enhanced noise or telegraph switching, and the
DCFs do not appear with highly reproducible energies.\cite{Csonka2004}


One possible interpretation of the data is that $dI/dV$ reflects a
convolution of the (thermally broadened) DOS of the source and drain
electrodes or a particle between the electrodes, as in scanning
tunneling spectroscopy.  In this situation the DCFs would correspond
to features in the DOS at energies $eV_{\mathrm{sd}}$ away from the Fermi level
($E_F$).  A probe of the electrode DOS would not be expected to produce
symmetric features with respect to $V_{\mathrm{sd}}$, nor would there be reason
for the peaks to shift towards $E_F$ with increasing $T$.
The symmetry of $dI/dV$ features with respect to $V_{\mathrm{sd}} = 0$,
together with the observed gate and temperature dependencies are also
inconsistent with expectations of Coulomb blockade physics
(\textit{e.g.}, resonant tunneling through a nanoparticle in
the interelectrode gap).

Pd can undergo a superconducting phase transition when doped with H
\cite{Skoskiewicz1972,Stritzker1972} or when irradiated with He ions
\cite{Stritzker1979} to create disorder and lattice strain.  The form
of the $T$-dependence of the DCFs in our devices is the same as that
used to describe the energy gap, $\Delta$, in superconducting
materials.  The $dI/dV$ data have some qualitative resemblance to that
seen in superconductor-superconductor tunnel junctions.  However, as
temperature is increased one would expect all $dI/dV$ peaks to
approach a common value of $T_{\mathrm{c}}$, contrary to the behavior of the DCFs
in our Pd break junctions.  Moreover, the gap energies inferred from
interpreting the DCFs as DOS features are orders of magnitude away
from what one would anticipate given the observed $T_{\mathrm{c}}$s.

Figure~\textbf{4} imposes a significant constraint on any candidate explanation
for the data.  It is difficult to construct a scenario in which the
energies of features in the DOS vary systematically with
interelectrode conductance as observed.  In a typical tunnel junction,
increasing the electrode separation will reduce the conductance
amplitude related to features in the DOS (\textit{e.g.} from
resonant tunneling into accessible states), but should not alter their
respective energies.  Fig.~\textbf{4} strongly suggests that the DCFs are not
simple DOS resonances.

\begin{figure}
\begin{center}
\includegraphics [scale = .33]{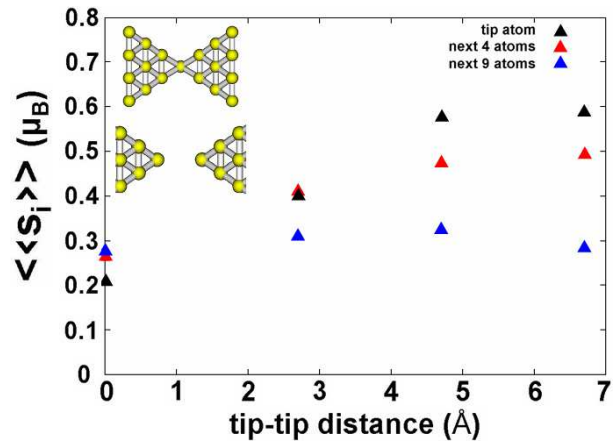}
\end{center}
\hspace\fill \vspace{-9mm} \caption{DFT calculations of magnetic moments for the end three layers forming the tip of a Pd electrode grown in the 001 direction as a function of the tip-tip distance, $d$.  Figures in the inset represent the model tips for contact (upper, $d$ = 0 \AA) and tunneling (lower) regimes.  Results show that Pd tips are magnetic, but with atomic magnetic moments smaller than those found in Pd chains (for the same type of calculation) due to higher coordination of the atoms.}
\label{DFTcalc1}
\end{figure}

The linear relationship between DCF $V_{\mathrm{sd}}$ and the inferred interelectrode separation $d$ suggests the possibility that the DCFs represent physics driven by the interelectrode electric field, $V_{\mathrm{sd}}/d$.  However, \textit{any} electric field-based mechanism is constrained by the mean field-like temperature dependence seen in Fig.~\textbf{3}.

One candidate mechanism would be electric field control of magnetic ordering, which has been suggested by several groups\cite{Duan2008,Ovchinnikov2009,Kudasov2007} and demonstrated experimentally in FePd.\cite{Weisheit2007}  Indeed, hole-doping of Pd with an electric field has been predicted to instigate a transition to ferromagnetism, causing a step-like change in the tunneling current at a given value of applied bias.\cite{Kudasov2007}  In our structures, the bias across the tunnel junction induces a local (on the scale of the screening length) accumulation of electrons and holes on the facing electrodes, resulting in a raising or lowering of the Fermi level at the surface, respectively.  Palladium has a peak in the DOS just below the Fermi level.  It is conceivable that a sufficient accumulation of holes could result in the satisfaction of the Stoner Criterion leading to itinerant ferromagnetism at that surface.

However, there are a number of problems with this scenario.  For example, this idea fails to account for the symmetry observed in the conductance spectrum about $V_{\mathrm{sd}} = 0 V$.  Which electrode is hole-accumulating depends on the bias polarity.  It seems unlikely that the two polarities would result in identical electric field-driven transitions.  It is also unclear how such a mechanism would lead to the rich variety of DCFs and their observed dependence on $V_{\mathrm{g}}$ (See Supplementary Information).  Furthermore, calculations indicate that the strength of the field necessary to instigate such a transition ($>$ 1 V/\AA) would require an applied bias prohibitively high in our devices.\cite{Kudasov2007,Sun2010}  Therefore, we do not think that electric field-driven magnetism is a likely explanation for our observations.

We speculate that the observed DCFs are related to inelastic electron
tunneling.  An \textit{inelastic} tunneling origin for the low bias DCFs would
naturally explain the observed symmetry in $V_{\mathrm{sd}}$, and the variety
of lineshapes could be explained by interference between various elastic and inelastic channels involving local modes.\cite{Mii2003,Galperin2004}  In contrast, the high bias DCFs are exclusively broad maxima, suggesting that different processes may be involved for low and high bias features.  However, the fact that both regimes are strongly correlated with one another (Fig.~\textbf{4}) and have the same trend with zero-bias conduction strongly implies that the low and high bias DCFs share crucial underlying physics.

We further propose that the relevant transport channels may arise due
to the onset of spontaneous magnetization at the electrode tips due to
the proximity of Pd to a magnetic instability.  While Pd surfaces are
not magnetic for film thicknesses exceeding a few monolayers, reports
suggest the possibility of itinerant ferromagnetism in Pd
nanoparticles,\cite{Lee1998,Sampedro2003} and atomic-size
contacts.\cite{Delin2004}  Zero-temperature density functional theory
(DFT) calculations for model Pd nanocontacts, as those shown in
Fig. \textbf{5}, indicate the possible presence of magnetism due to
the low coordination of the atoms forming the electrode tips. (See
Supplementary Information for calculation details).  A quantitative
estimate of the Curie temperature is, however, beyond DFT
capabilities.  Furthermore, even if we take the calculated energy
difference per atom between the ferromagnetic and paramagnetic states
as a reference value, the inherent uncertainties in the calculations
are larger than the range of the observed DCF $T_{\mathrm{c}}$s.
Nevertheless, we can say that this energy difference sets an upper
limit to $T_{\mathrm{c}}$ that is consistent with the observed values
(see Supplementary Information).

If one assumes an excitation energy proportional to the local
magnetization (consistent with, \textit{e.g.}, Stoner excitations), then the
temperature evolution of $eV_{\mathrm{sd}}$ would correspond to the
onset of magnetic order.  Moreover, within that assumption the same
model calculations are consistent with the trend of Fig.~\textbf{4}.
The layer-by-layer magnetic moment per atom (averaged over layer),
$<\langle{s_i}\rangle>$, is plotted in Fig. \textbf{5} for the three
atomic end layers of one tip (the other tip is identical) as a
function of the distance between tips, $d$.  The calculated average
magnetic moment per atom decreases as one moves away from the ends of
the tips towards the bulk, revealing the expected inhomogeneous nature
of the magnetization.  The magnetic moment of the tip atom is visibly
affected by the interelectrode distance, decreasing with $d$ below 5
\AA, consistent with the dynamic range of $d$ estimated earlier from the measured ZBC. This can be rationalized by noting that on increasing the coordination (decreasing $d$) the DOS decreases at the Fermi level and the polarization becomes weaker, in accord with the Stoner criterion.
The electrodes in a real break junction may contain many
undercoordinated atoms, giving rise to localized areas of increased
magnetic susceptibility, extending over a larger region of the
junction. In this picture, DCFs with different $T_{\mathrm{c}}$s would
result from different local portions of the electrodes.


Another related speculative mechanism for DCFs involves the interplay
of spontaneous magnetization, as above, and Kondo physics.  The
formation of the highly correlated Kondo state in bare metal break
junction devices\cite{Ralph1992,Houck2005} can lead to sharp features
in $dI/dV$ at symmetric voltages.  In this case the $dI/dV$ peaks
would correspond to a spin-split zero-bias resonance resulting from a
strong local exchange field established by the presence of
ferromagnetism\cite{Pasupathy2004,Calvo2009} or effective magnetic
impurities.\cite{Heersche2006}  The $T$-dependent onset of local
magnetization at the electrode tips could lead to such a splitting
that increases with decreasing temperature.

The apparent lack of magnetic hysteresis and significant
magnetoresistive effects would seem to cast doubt on a hypothesis
involving magnetic correlations.  However, it is worth recalling that
Stoner-like excitations would not Zeeman split, although they may be
expected to shift with the application of an external magnetic field.
Furthermore, magnetically ordered regions possibly present very small
local anisotropy energies\cite{Wierzbowska2005} and a
superparamagnetic response to externally applied fields.  This implies
a vanishingly small coercivity and suggests that large hysteretic
effects are not necessarily expected.\cite{Skumryev2003}  The mismatch
between DCF energy scale and $T_{\mathrm{c}}$ is not readily
explained, and would require a more detailed theoretical picture of
the inferred magnetic transition and resulting excitations.  One
should recall, however, that Stoner excitations (invoked as a possible
explanation of DCFs) and the spin stiffness (related to
$T_{\mathrm{c}}$) are associated with different exchange
couplings.\cite{Pajda2001}  This makes their characteristic energy
scales differ by an order of magnitude, at least in transition metals.
We see no reason to expect otherwise in the case of Pd.


The data presented here unambiguously show that unusual electronic
effects are present in Pd nanojunctions at low temperatures.  The
trends in the data with $T$, $G(V_{\mathrm{sd}}=0)$,
$V_{\mathrm{sd}}$, and $V_{\mathrm{g}}$ greatly restrict possible
mechanisms.  While speculative, an explanation of the features in
terms of inelastic processes involving magnetism is consistent
with the observed temperature range, source-drain symmetry, special
character of Pd, and the trend with inferred electrode separation.
This idea is also testable through doping the Pd with a ferromagnetic
metal to see if enhancing the tendency toward ferromagnetism alters
the phenomenon.  These conductance anomalies demonstrate once again
that even ``simple'' chemically homogeneous materials can exhibit rich
and surprising phenomena at the nanoscale.

\textbf{Methods}\
Arrays of bowtie-shaped nanowire patterns are defined by electron beam lithography with minimum widths between 80 and 100~nm.  Samples are composed of 14 nm of Pd deposited by e-beam evaporation on a highly doped Si substrate with a 200~nm SiO$_2$ insulating layer (Fig.~\textbf{1}a).  The devices are contacted individually using a piezo-controlled Attocube\texttrademark probe station inside of a $^4$He cryostat.  An array of devices is cleaned in solvents, exposed to an oxygen or argon plasma clean, and subsequently inserted into the sample space of the probe station, which contains approximately 10~mB of He exchange gas at 300~K. The electromigration technique\cite{HPark1999} is employed at low temperature ($T \lesssim$ 2~K) to create a small gap in the nanowire (Inset, Fig.~\textbf{1}a).  Differential conductance ($dI/dV$) is measured as a function of source-drain bias ($V_{\mathrm{sd}}$) using standard low frequency lock-in techniques with an rms excitation voltage of 625~${\mu}$V.  I-V is also measured using a voltage source and current preamplifier, for a small portion of samples.

\textit{Acknowledgment.} D.N. acknowledges support from the David and Lucille Packard Foundation and the W. M. Keck Foundation.  G.D.S. acknowledges the support of the W. M. Keck Program in Quantum Materials at Rice University.  The authors would like to thank Joaqu{\'i}n Fern{\'a}ndez-Rossier, David Jacob, Qimiao Si, Jan van Ruitenbeek, Christian Sch{\"o}nenberger, Michel Calame, Nadya Mason, Allen Goldman, and Laura Greene for comments and discussions, Dan Ward for assistance with SEM images, and Emilia Morosan and Liang Zhao for SQUID measurements.

\textit{Supplementary Information Available.} This includes additional details with regard to experimental setup and results not included in the main text.  The effect of an applied gate bias is addressed as well as results pertaining to tests performed in the presence of hydrogen gas.  Lastly we address the details of the DFT calculations.

\end{document}